\documentclass[useAMS,usenatbib]{mn2e}
\usepackage{graphics,epsfig,psfig}
\usepackage[normalem]{ulem}
\usepackage[dvipsnames]{color}
\usepackage[]{inputenc,amssymb}
\usepackage{ulem}

\def \be{\begin{equation}}
\def \ee{\end{equation}}
\def \bea{\begin{eqnarray}}
\def \eea{\end{eqnarray}}
\def\etal{{et al.\ }}
\def\ltsima{$\; \buildrel < \over \sim \;$}
\def\simlt{\lower.5ex\hbox{\ltsima}}
\def\gtsima{$\; \buildrel > \over \sim \;$}
\def\simgt{\lower.5ex\hbox{\gtsima}}
\def\zsun{Z_\odot}

\title[Evolution of multiple supernova remnants]{
Evolution of multiple supernova remnants}

\author[Evgenii O. Vasiliev, Biman B. Nath, Yuri Shchekinov]
{Evgenii O. Vasiliev$^{1,2}$,  Biman B. Nath$^3$, Yuri Shchekinov$^{1,4}$\\
$^1$Department of Physics, Southern Federal University, Rostov on Don 344090, Russia\\
$^2$ Institute of Physics, Southern Federal University, Rostov on Don 344090, Russia\\
$^3$Raman Research Institute, Sadashiva Nagar, Bangalore 560080, India\\
$^4$Isaac Newton Institute of Chile, SAO RAS Branch, Russia\\
}

\begin{document}


\maketitle

\label{firstpage}

\begin{abstract}
Heating of the interstellar medium by multiple supernovae (SNe) explosions is at the heart of 
producing galaxy-scale outflows in starburst galaxies. Standard models of outflows assume a 
high efficiency of SNe in heating the gas to X-ray emitting temperatures and filling 
the central region of starburst with hot gas, in order to launch vigorous outflows. 
We use hydrodynamical simulations to study the efficiency of multiple SNe in heating the interstellar medium (ISM) and filling 
the volume with gas of high temperatures. 
We argue that it is important for SNe remnants to have a large filling factor {\it and} a large
heating efficiency. For this, 
they have to be clustered in space and time, and keep exploding until
the hot gas percolates through the whole region, 
in order to compensate for the radiative loss.  
In the case of a limited number of SNe, 
 we find that although the filling factor can be large, the heating efficiency declines after reaching
a large value. 
In the case of a continuous series of SNe, the hot gas ($T \ge 3 \times 10^6$ K) can
percolate through the whole region after the total volume filling factor reaches a threshold of $\sim 0.3$. The efficiency of heating
the gas to X-ray temperatures can be $\ge 0.1$ after this percolation epoch, which occurs after a period of
$\approx 10$ Myr for a typical starburst SNe rate density of $\nu_{\rm SN} \approx 10^{-9}$ pc$^{-3}$ yr$^{-1}$ and gas density of $n\approx 10$ cm$^{-3}$
in starburst nuclei regions. 
This matches 
the recent observations of a time delay of similar order between the onset of star formation and galactic outflows.
The efficiency to heat gas up to X-ray temperatures ($\ge 10^{6.5}$ K) roughly scales as 
$\nu_{\rm SN}^{0.2} n^{-0.6}$. For a typical SNe rate density and gas density in starburst nuclei, 
the heating efficiency is  $\sim 0.15$,
also consistent with previous interpretations from X-ray observations.
We discuss the implications of our results with regard to observational diagnostics of ionic ratios and 
emission measures in starburst nuclei regions.
\end{abstract}


\section{Introduction}
Supernovae provide a major source of feedback in the ISM  of galaxies.    
The energy released by SNe, about $10^{51}$ erg in kinetic energy, is deposited in the ISM through the action of the blast waves driven by them. The total volume of the ISM engulfed by the SN remnant and the rate at which the energy is dissipated depends on the density and temperature of the ambient medium. These aspects have been studied with analytic calculations and numerical simulations in the literature \citep[e.g., ][]{cox72, chevalier74, cioffi88, thornton98, shelton98}. However these studies have focused on the evolution of isolated SNe. Core-collapse SNe are related to the end stages in the evolution of massive stars (OB), and they are  mostly clustered because massive stars  form in OB associations. Therefore the SNe events are likely to spatially and temporally clustered. The evolution of multiple SNe overlapping with one another in time and space is likely very different from that of isolated SN remnants because of the different ambient conditions encountered by them. 

The concerted effect of clustered SNe is believed to a superbubble \citep[e.g., ][]{maclow88, koo92}, whose shell of swept up mass moves faster than the typical speed of OB associations (few km s$^{-1}$), and which therefore contains most of the SNe arising from the association. The study of the evolution of these superbubbles has mostly assumed continuous energy release from the centre. Tang \& Wang (2005) have studied the effect of sporadic SNe explosions occurring in the low density medium inside superbubbles. They found that SNe shells move in the hot, low density medium faster than predicted by Sedov-Taylor solution, and suffer less radiative loss. This problem has also been recently studied in detail by \citet{sharma14}. However, the SNe explosions in their simulations were assumed to occur at the same location. It remains to be seen how clustered--- although not necessarily spatially coincident--- SNe events affect the surrounding medium.

This problem becomes acute in the context of supernovae driven galactic winds in which it 
is assumed that SNe can sufficiently heat up the ISM gas, at least in the central region of disc
galaxies, in order to launch a wind. This process assumes that although SNe lose most energy in radiation in isolated cases, the efficiency of heating the ISM can be large in the central region filled with hot and low density gas and that the gas in this region is thermalized \citep[e.g., ][]{clegg85, sharma13}.  Numerical simulations \citep[e.g., ][]{suchkov94, suchkov96, strickland00, fujita09, strickland09} also implement the initial conditions leading to galactic winds making similar assumptions. It is believed that in a multiphase medium and in the case of multiple 
SNe events, the efficiency of SNe heating --  the fraction of the total explosion energy transferred into thermal 
energy --  can be larger than $\sim 0.1$. 
These estimates  came from the numerical and analytical studies of energy loss in {\it isolated} supernova remnants, which showed that the fractional energy retained in the hot interior gas of remnants was of order $\sim 0.1$. 
\citet{larson74} had first pointed out the importance of cooling with regard to galactic outflows, and derived a critical supernova rate density required to compensate for cooling. His estimate was based on the results of single SNR evolution by 
\citet{chevalier74b} and \citet{cox72}, which stated that thermal energy retained by a SNR is of order $20\%$ at the time when radiative losses begin to dominate (say, at $4t_r$). These results were verified later by  detailed simulations of 
\citet{thornton98} and further showed that the fraction steadily decreases to about $\sim 3 \%$ after a time scale 
of $10 t_r$. Our goal is to extend these estimates to the case of multiple SNe.

The question of heating efficiency of SNe crucially depends on the evolution of multiple SNe which has not yet been studied in detail. With the advent of X-ray studies of galactic outflows, the problem has become more pressing because {\it one not only has to find the conditions for high efficiency of heating by supernovae, but also for a large filling factor for X-ray emitting gas with $T\ge 10^6$ K.}
 Although Melioli \& de Gouveia Dal Pino (2004) studied the average heating efficiency of multiple SNe, the filling factor of hot gas ($\ge 10^6$ K) was not estimated. Recently
\citet{nath13} have argued that the energy input from multiple SNe in the central
regions of starbursts cannot heat the gas to $T \ge 10^6$ K, unless the SNe events  act coherently
in space and time.  More precisely, coherent action suggests  that successive SNe shockwaves mostly 
propagate into a hot medium, which has been heated by earlier SNe and which has not had time to radiatively cool. This coherency condition is ensured when the SNe shells collide with one another before they they enter the 
radiative (pressure-driven) phase and lose most of their energy.  However, \citet{nath13} did not consider the long term behaviour
of multiple SNe when nonlinear effects change overall dynamics and distort simple order of magnitude estimates. One of such nonlinear 
behaviour is the possibility of the most hot inner gas to percolate throughout the multi-temperature starburst region.

 To summarise, two conditions are essential in order to excite outflows: (1) the SNRs should fill a substantial fraction of the total
volume of the star forming region, and (2) the efficiency of heating the gas to high temperature ($\ge 3 \times 10^6$K) should
be large ($\ge 0.1$). The first suggests that all SNe act collectively and inject their energy basically into the whole mass of star forming region. The second implies that 
a certain fraction of the injected energy is unavoidably lost through radiative processes, but  a non-negligible still remains for producing the gas outflow.
 It is often assumed that the first condition automatically leads to the second condition. For example, 
\citet{heckman90} used the filling factor argument from \citet{mckee77} (which was originally done for a lower temperature gas, at
$3 \times 10^5$ K), and assumed that it would lead to large heating efficiency. \citet{clegg85} assumed a comparably large filling factor
and heating efficiency for SNRs in the nucleus of M82.

As fas as the large scale effect of multiple SNe associated with a starburst -- the galactic wind -- is concerned, an additional and physically independent condition is normally applied. 
The explosion energy or the mechanical luminosity is sufficient to 
break through the gaseous disk where the starburst has occurred and then to launch the residual gas mass away \citep{ks85,maclow88,ft00}. In numerical simulations 
the mechanical luminosity of a central engine is always assumed to satisfy the third condition, while the former two are implicitly supposed to have fulfilled. Even though 
these three conditions can at certain circumstances overlap in the parameter space they are physically independent.  {\it We therefore emphasise in this paper that these conditions should be separately met,
and this requirement puts a more realistic threshold on the SFR.} Also, this leads to a prediction for the time lag between the
onset of star formation and the launching of outflows, which is shown to be consistent with observations.

In this paper, we  therefore study two aspects of the problem of multiple SNe, namely, to test the importance of coherency condition, and
to study the time scale and conditions under which percolation of hot gas becomes possible, with hydrodynamical simulations. 
This allows us to study the efficiency of heating by multiple SNe events, in particular the efficiency of heating gas to high temperature. 
We then discuss the implications of the low filling factor of hot gas on the onset of galactic winds, and some observable properties of warm and hot gas.

\section{Theoretical estimates}
Consider the case of multiple SNe that explode simultaneously, after which their remnants interact with
each other.
In the Sedov-Taylor phase of a single SN remnant evolution is described by the scaling $R\sim C (E t^2/\rho)^{1/5} $, where $E$ is the explosion energy,  $\rho$ is the mass density, and the constant $C = 1.15$. If the number  of SNe explosion is denotes by $N$, and the computational domain is of size $R_{\rm comp}$,  or, if $n_{\rm SN}$ is the number density of SNe,
 then the porosity
is defined as the fraction of volume occupied by all SNe remnants as if they were isolated,
\be
Q= {N \over R_{\rm comp}^3} 
{4 \pi \over3} R(t)^3 = n_{\rm SN} 
\, C {4 \pi \over 3} \Bigl ({E t^2 \over \rho} \Bigr )^{3/5} \,,  
\label{eq:por}
\ee
This definition of porosity is essentially statistical in nature, in the sense that this assumes an ensemble of large number of computational zones, each containing several SNe explosions spread randomly in space. In other words, this definition neglects the possible effects of spatial clustering of individual SNe.

The porosity keeps increasing as long as the shocks are strong. \citet{mckee77} showed that in the case of a steady rate of SNe explosions ($\nu_{SN}=$ const), and with $R \propto t^\alpha$, the porosity is given by $Q(t)=(4 \pi/3)  \nu_{SN} R^{3} t (1+3 \alpha)^{-1}$.
The filling factor for a random distribution of SNe explosions is given by $f=1-\exp (-Q)$ \citep{mckee77}. Since $R \propto t^{2/5}$, we have $Q \propto t^{6/5}$; initially $f$ increases almost linearly with time, but later if the remnants begin to interact and merge with each other, the subsequent evolution of $f$ becomes a weak function of time. 
Note that the filling factor here refers to gas inside SN remnants, consisting of gas at different temperatures. 

In the adiabatic case, the filling factor of the hot ($\ge 10^6$ K) gas first increases, as the individual SNe remnants expand and shock heats the engulfed gas. At a later stage, as the shell speed decreases, the gas is heated to a lower temperature. The hot gas at this stage expands 
behind the shock front and cools adiabatically, thereby decreasing the filling factor of the hot gas. In contrast, the filling factor of warm ($10^5$ K) gas asymptotically increases, and becomes close to the overall filling factor of the shells, because most of the shell volume gets filled with this gas. 

The inclusion of radiative loss changes the evolution of the filling factors of gas with different temperature in the following manner. Firstly,
the average temperature of the gas that fills the remnant volume depends on the age, explosion energy and the ambient density, and it decreases rapidly after reaching $10^6$ K because of enhanced cooling below this temperature. Cox (1972) showed that the total energy of the remnant scales as $E(t) \propto R^{-2}$, after the shell enters the radiative phase (at shell radius $R_c$), and the total energy drops to half the initial value, when the average temperature of the inner
gas decreases to $\sim 10^6$ K. \citet{chevalier74b} also derived a similar result, that $R \propto t^{0.31}$. 
This is supported by simulations \citep{shelton98, thornton98}. The gas with very high temperature does not lose energy through radiation, as it is mostly low density gas. It is the lower temperature gas that cools rapidly after this stage.

In the radiative case, one therefore expects the filling factor of $10^6$ K gas to first increase as individual SNe remnants expand until the radiative phase, as in the adiabatic case. 
The subsequent evolution of the hot gas depends on whether or not the SNe events occur in a coherent manner. The coherency condition was defined in Roy \etal (2013), Nath \& Shchekinov (2013) in such a way that SNe events occur continuously with a sufficient rate density (per unit time and volume) in order to compensate for the radiative loss. If the shell radius at the radiative epoch $t_r$ (when loss due to radiation becomes important) is given by $R_0$, and if the steady state rate density of SNe events is denoted by $\nu_{\rm SN}$, then the condition for coherency is for the four-volume, $\nu_{\rm SN} \times V(t_r) \times  t_r   \ge 1$, where $V(t)$ is the volume of a single SN remnant at time $t$. This condition essentially implies that the SNe remnants are coherent when they overlap before cooling radiatively.  

In order to act together, the multiple SN remnants need to overlap. The time scale over which SN remnants overlap ($t_c$) can be defined as the time when the computational box is occupied by shells. We therefore have, 
\be
n_{\rm SN} 
V(t_c) 
\approx 1 \,.
\label{eq:tcol}
\ee
This time scale $t_c$ marks the transition between isolated SN remnants to a collective bubble.

However, in order for their combined energy to be effective without suffering much loss, the overlap should occur before the end of the adiabatic Sedov-Taylor phase, or the beginning of the
radiative phase. 
The radiative loss time scale can be estimated by requiring the shell speed to drop to $\sim 120$ km s$^{-1}$, so that the post-shock temperature $T_{\rm sh} \sim 2 \times 10^5$ K (for $\mu=0.6$),
where the cooling function peaks. This gives a time scale and shell radius,
\be
 t_r =0.14  \Bigl ( {E_{51} \over n} \Bigr ) ^{1/3} {\rm Myr} \,;  R (t_r) =37  \Bigl ( {E_{51} \over n} \Bigr )^{1/3} \, {\rm pc} \,, 
\label{eq:trad}
\ee
 where $E=10^{51} \, E_{51} $ erg,  $n$ is particle density in cm$^{-3}$.
The peak of the cooling function depends weakly on metallicity, and so $t_r$ is roughly independent of metallicity. 
This is also close to the time scale when half of its thermal energy is radiated away, $\sim 0.17$ Myr $n^{-1/2}$, for
standard cooling function \citep[eq. 7 of ][]{babul92}.

Therefore, for a computational box of size $R_{\rm comp}$ and $N$ number of SNe going off within the time scale $t_r$, we have the  condition, 
\be
{n_{\rm SN}  \over 
t_r} \times ({4 \pi \over 3})  R (t_r) ^3  \times t_r \ge 1 \,.
\label{eq:coh1}
\ee

In other words, the above conditions boil down to $V(t_c) \le V(t_r)$, or equivalently, $t_c\le t_r$. This is therefore the coherency condition, which states that the volumes of SN remnants merge before they radiate away most of their energy, 
thus ensuring the energy released by individual SNe to act in a concerted manner. 
Note that the radiative loss may be enhanced to some extent before $t_r$ even in the coherent case due to compressed gas in interfaces of merging shells.

It is, however, not enough to simply fill the star forming region with SN remnants, as temperature and density distributions in their interior are non-uniform and have different rates of radiation loss. It is important to ensure that the explosion energy from SNe propagates 
mostly  into a hot and tenuous medium for the heating efficiency to be large. This requires a certain period of time for the SNe remnants to fill a substantial fraction of the relevant volume, when the inner hot ($T\geq 3\times 10^6$ K) gas would be able to percolate freely. This will ensure that later generation of SNe would explode in an already hot medium and thus lose less energy. Therefore it is crucial that star formation should continue until the percolation takes place.

The possibility
that hot gas would percolate after a sufficient fraction of the total star forming region is engulfed,
 has been postulated since the original paper of \citet{larson74}, but not been demonstrated. One requires hydrodynamical simulations
 to find out the threshold filling factor of the SNe remnants after which the hot gas can effectively fill up the volume and heating efficiency
 can become large. It is important to first establish the phenomenon of percolation in this regard, and to determine the relevant time
 scales as a function of various parameters, such as SNe rate density and the gas density.

Although these ideas are both intuitive and straight-forward, their implications have not been discussed in detail in literature. This is compounded by the lack of knowledge of how rapidly energy is lost in the multiple SNe case. As we will show below, the energy loss rate in the cases of multiple SNe and isolated remnants {\it are different}. The standard practise in the literature has been to either use the energy loss rate of isolated remnants, or to hope that somehow multiple SNe would fill the region with dilute gas which would not radiate much. It is assumed that a central region of $\sim 200$ pc is filled with hot gas that has been heated with an efficiency of order of unity \citet{clegg85}. 
The original motivation for this assumption using the arguments of \citet{mckee77} refers only to gas with $T \sim 10^{5.5}$ K and not the X-ray emitting gas, as has been shown by \citet{nath13}. 
As we will show below, this expectation ignores the effect of dense gas accumulated in the merged shells. In a recent paper, \citet{sharma14} have shown that a heating efficiency of $\sim0.3$ is achievable only in the case of a high degree of spatial and temporal coherence, namely in the case of coincident SNe separated by short time intervals, in the case of a superbubbles driven by OB associations. They have shown the standard assumption of a thermal wind solution of \citet{clegg85} is valid only in superbubbles with short intervals between SNe. Our simulations generalise these results to the case of multiple SNe that are spatially (and temporally) separated.


\citet{melioli04} studied the effect of multiple SNe in heating the ISM, and estimated a heating
efficiency, averaged over gas with different temperatures, in different situations. They found that 
the heating efficiency is initially small, in the range $0.01\hbox{--}0.1$, and it rises sharply after
$\sim 20$ Myr after the onset of star formation, when the most of the gas has been expelled from the region and the gas density has become small. However, it is not clear how this time scale depends on ISM parameters such as density, metallicity and
the SFR.
Another recent simulation of SNe driven wind by \citet{stringer12} 
They 
determined the fraction of energy that is converted into thermal energy to be $\sim 0.024$ (their Figure 2 and footnote 9). 
\citet{creasy13} found the fraction to lie in the range $0.03\hbox{--}0.4$, depending on the  gas mass in the galaxy (see their Figure 11). The lower the  gas fraction, the more tenuous is the gas and the higher is the heating efficiency. Interestingly, the commonly used value of $\sim 0.3$ for the heating efficiency is the {\it least conservative estimate} in all these simulation results. Apart from these, 
\citet{hill12} studied the vertical structure of a magnetised ISM in a thermally driven outflow, and found a filling factor of $\sim 0.2$ for gas above $T \ge 10^{5.5}$ K. \citet{hopkins12} have studied the feedback process in the ISM of disc galaxies because of SNe driven winds, but their focus was on the relation between mass loss rate and the star formation rate, and not on the energetics of SNe driven winds. In summary, the recent simulations study the whole process of feedback and it is difficult to disentangle the heating efficiency from the effect of other parameters (since they are all linked), and the  oft-quoted values of efficiency either relies on analytical estimates based on the earlier results of single SN remnants, or use the least conservative values from numerical simulations mentioned above.

\section{Numerical method and initial conditions}
We use three-dimensional unsplit total variation diminishing (TVD) code based on the Monotonic Upstream-Centered Scheme for Conservation Laws (MUSCL)-Hancock scheme and the Haarten-Lax-van Leer-Contact (HLLC) method (e.g. Toro 1999)
as approximate Riemann solver. This code has successfully passed the whole set of tests proposed in (Klingenberg 
\etal 2007).

In the energy equation we take into account cooling processes adopted the tabulated non-equilibrium cooling curve 
(Vasiliev 2013). This cooling rate is obtained for a gas cooled isobarically from $10^8$ down to 10~K. Given the typical temperature and sound speed considered in our problem, the sound crossing time of a resolution element in our simulation is of order $\sim 1000$ yr, much shorter than  the relevant time scales in the problem, and therefore we choose isobaric cooling. The
non-equilibrium calculation includes the ionization kinetics of all ionization states for the following chemical 
elements H, He, C, N, O, Ne, Mg, Si, Fe as well as molecular hydrogen kinetics at $T<10^4$~K (see Appendix 1 for details). 
The tabulated ionization states are used for calculating column densities and emission measure. The heating rate is
adopted to be constant, whose value is chosen so that the background gas does not cool. 
The stabilization vanishes when the density and temperature goes out of the narrow range near the equilibrium state.

We have carried out 3-D hydrodynamic simulations (Cartesian geometry) of multiple SNe explosions. We consider periodic boundary conditions. The computational domains have size $200^3$ pc$^3$, which have $300^3$ cells, corresponding to a physical cell size of $0.75$ pc. 
The background number density considered range between $0.1\hbox{--}10$ cm$^{-3}$,
and the background temperature is $10^4$~K. The metallicity is constant within the computational
domain (we do not consider here the mixing of metals ejected by SNe, this question will be studied in a separate
work), and we consider cases with $Z=0.1, 1$ Z$_\odot$. 
We inject the energy of each SN in the form of thermal energy in a region of radius $r_i=1.5$ pc. 
SNe are distributed uniformly and randomly over the computational 
domain. 

\begin{figure}
\centerline{
\epsfxsize=0.5\textwidth
\epsfbox{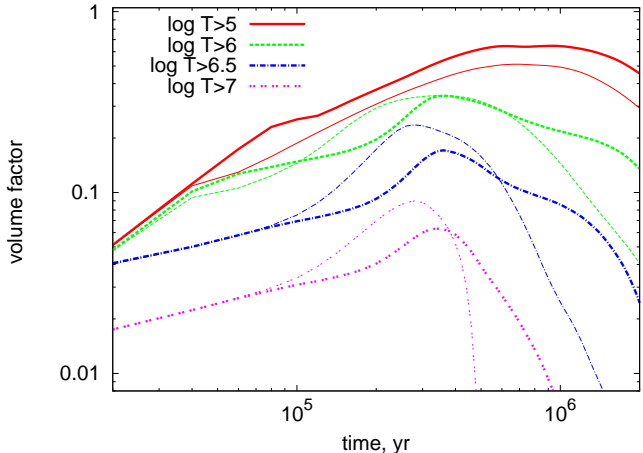}
}
{\vskip-3mm}
\caption{
Filling factors of gas with different temperatures for simultaneous SNe explosions. Gas number density is 1~cm$^{-3}$, with $0.1$ Z$\odot$  (thick lines) and solar metallicity (thin lines).
The lines correspond to  gas with
$\log~T>$ 5, 6, 6.5, 7 (from top to bottom). 
}
\label{fig:3dsim}
\end{figure}

In the following, we will begin by studying  how the coherency condition is related to the filling factor of hot gas in the short run. We will then 
continue to
study the long term behaviour of a large number of SNe remnants, and the implications on the heating efficiency and subsequent
launching of a galactic outflow.

\section{Results}

\subsection{Simultaneous SNe explosions}
We first study the effect of 15 SNe exploding simultaneously in a computational domain of size $200^3$ pc$^3$. 
The density of SNe is $n_{\rm SN}\approx 1.9 \times 10^{-6}$ pc$^{-3}$ and they are distributed randomly in the computational domain.
The gas particle density is $1$ cm$^{-3}$ with metallicity $Z=0.1$ Z$_\odot$ and solar metallicity in two different models, and the explosions have energy $10^{51}$ erg. The timescale for overlap of SNe is  $t_c\sim 0.16$ Myr, from eqn \ref{eq:coh1}. 
For $n=1$ cm$^{-3}$, the radiation loss time scale $t_r \sim 0.15 $ Myr. Therefore one has $t_c \sim t_r$, and the coherency condition is marginally satisfied.  As mentioned earlier, these time scales weakly depend on metallicity. We show in Fig \ref{fig:3dsim} the volume
filling factor as a function of time elapsed, for $0.1$ Z$_\odot$ (thick lines) and solar metallicity (thin lines). The curves show that the filling factor of gas reaches a maximum at $t_c$ after which it decreases.
In particular, the filling factor of gas with $T\ge 3 \times 10^6$ K reaches $\le 0.2$ after $t_c$, after which it
rapidly decreases.



\begin{figure}
\centerline{
\epsfxsize=0.5\textwidth
\epsfbox{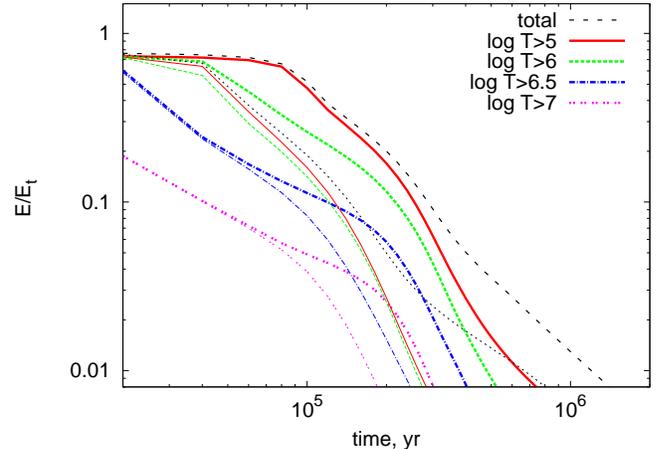}
}
{\vskip-3mm}
\caption{
The evolution of the ratio of thermal energy to the total energy, for gas in different temperature range is plotted for $Z=0.1 Z_\odot$ (thick lines) and solar metallicity (thin lines), for the case of 
simultaneous explosions.}
\label{fig:3deff}
\end{figure}

\begin{figure*}
\includegraphics[width=8.8cm]{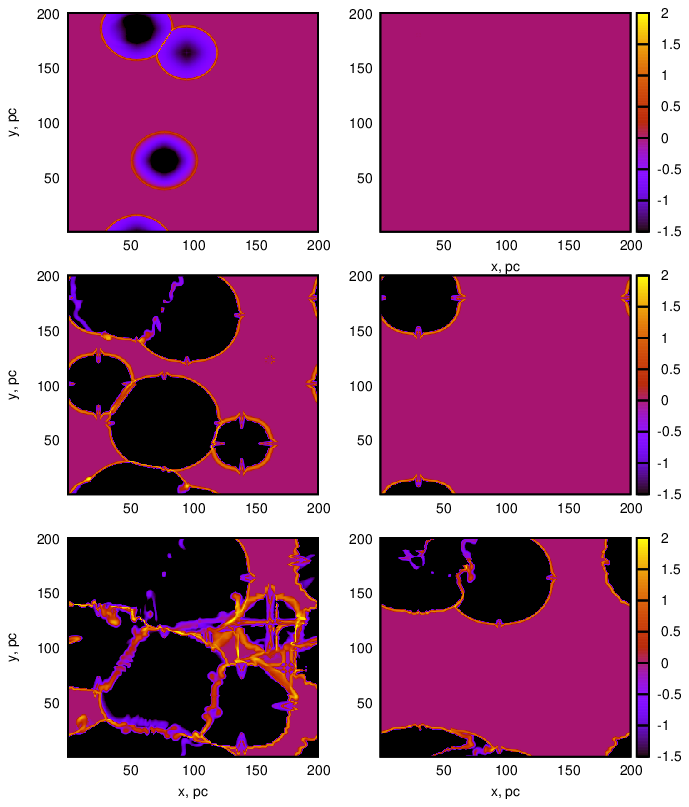}
\includegraphics[width=8.8cm]{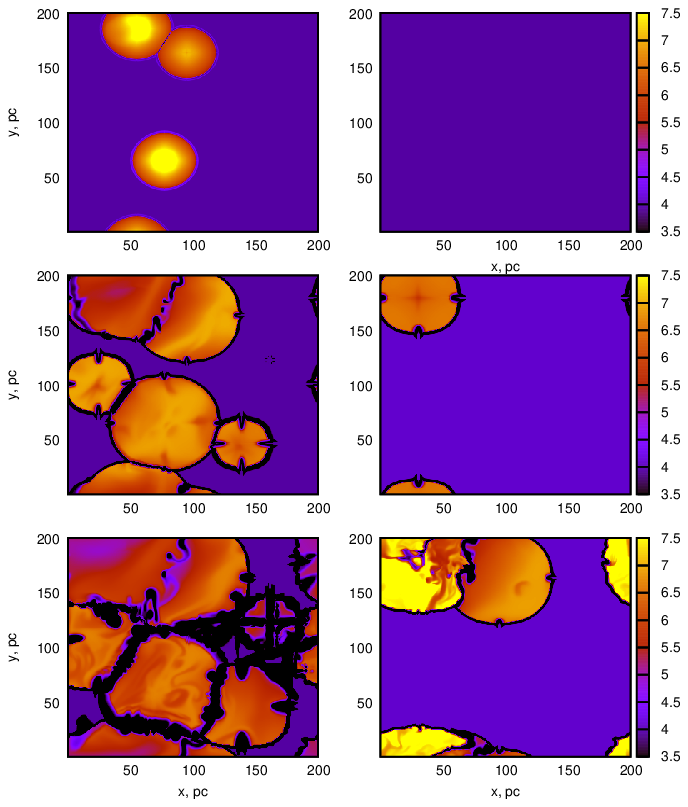}
\caption{
Density (left) and temperature (right) contours for snapshots at three different epochs (from top to bottom), at
$10^5, 5 \times 10^5, 10^6$ yr. The left column refers to the case with a delay of $10^4$ yr and the right column
refers to a delay of $5 \times 10^5$ yr, corresponding to dotted and dashed lines in Figure \ref{fig:3d-delay}, respectively.
The slices shown here are in $xy$ plane at $z=100$ pc, in the middle of the computational box.
}
\label{fig:contours}
\end{figure*}

Figure \ref{fig:3deff} shows the corresponding evolution of thermal energy stored in gas with different temperatures, for solar (thin lines) and $Z=0.1 \, Z_\odot$ metallcity (thick lines), for the same gas density and density of explosions as in Fig \ref{fig:3dsim}. The curves are normalized to the initial energy, and therefore show the fraction of energy of gas with different temperatures.The black lines show the total heating efficiency of gas, or the relative thermal energy stored in gas compared to the initial energy.  The fractional energy for gas at high temperatures ($\ge 3 \times 10^6$ K) reaches a value of $\sim 0.1$ around a time scale of $t_r$, and rapidly declines
afterwards. This makes the heating efficiency decline although the SNe are coherent in time (i.e., the remnants merge before radiating much of their energy) and the remnants can fill up a large fraction of the volume given
sufficient time.

The curves in Fig \ref{fig:3deff}  also show that after the injection from explosions is turned off, 
the total thermal energy evolves with time roughly as $\propto t^{-1.3}$,  
reasonably close to the scaling $\sim t^{-1}$ obtained  by 
\citet{thornton98} in the case of single SN remnants. The energy contained in hot gas ($T \ge 10^{5}$ K) scales 
with a steep function of time, as $t ^{-2.3}$ after the injection from explosions turns off. 
Therefore, the radiative loss in the case of multiple SNe is prohibitively large, owing to the large densities produced 
during the merging of shells and the consequent enhanced cooling.

\subsection{SNe explosions with time delay}
Next, we study SNe explosions with time delay, with two different delay periods, 
$\Delta t=10^4$ and  $5 \times 10^5$ yr. 
 The staggered sequence of SNe in each case are 1,2,3,4,3,2,1 after $\Delta t$, so that the SNe are over in each case after $7 \times \Delta t$. E.g, for $\Delta t=10^4 $ yr, all the SNe occur in the above staggered sequence within $7 \times 10^4$ yr. We can define the average time lag, averaged over the staggered sequence, to be the total time elapsed divided by the 
total number of SNe. E.g., for the time lag $\Delta t=10^4$ yr, we have the average time lag $\langle \Delta t \rangle=0.4 \times 10^4$ yr.
Therefore, in the first case the average time lag is less than the radiative time scale $t_r (\sim 0.15$ Myr), and in the second case, the average interval ($0.2$ Myr)  is also comparable to $t_r$. In other words, both cases satisfy the coherency condition.

We show the density and temperature contours of a few snapshots for the two cases in Figure \ref{fig:contours}, in the left and right panels, respectively. In each panels, the left column shows the case for $\Delta t=10^4$ yr and the right column, for $\Delta t=5 \times 10^5$ yr. The snapshots correspond to $t=10^5, 5\times 10^5, 10^6$ yr, from top to bottom. The temperature contours show that for the short time delay case, all the SNe have exploded by $10^5$ yr (the first snapshot), but most of their remnants have not yet entered the radiative phase. The subsequent snapshots show the effect of adiabatic and radiative energy loss. They show a growing dominance
of high density regions in the shells around merged remnants. 
The fact that {\it merged} shells give rise to high density regions, more than in the case of isolated remnants, has an interesting implication. These high density regions make radiative loss a bigger drain of energy in the case of multiple SNe than isolated supernova explosions. 
 The snapshot at $10^6$ yr shows the lack of very hot gas for the shorter time delay case. There are therefore three discernible stages in multiple SNe: adiabatic, isolated radiative and merging radiative, the last phase being an additional drain of energy compared to the case of isolated SNe.
For the longer time delay case, a similar progression from adiabatic to isolated radiative and further to merging radiative phases occur, although over a longer time scale. 

Note that there are some contaminations arising from the numerical strong-shock instability also referred to as the odd-even instability described by \citet{qk94}. 
This instability arises when shock fronts propagate along a grid axis and manifests in the growth of a bump on a front parallel to a grid plane. 
Radiation energy losses enhance the instability \citep{suth03}, and in our case it results in formation of spikes inside the shell which then develop 
cross-like artificial structures within bubbles. The spikes become visible immediately after the shell enters radiation phase, grow over a next couple 
of radiation time scales and may form crossing planes parallel to grid planes, and then finally disappear by 3 radiation time scales. 
The cross-like artifacts are therefore most pronounced at when 
the remnant is within 2-3 radiation time scales after entering the radiation phase: see, for instance, the youngest remnant on the bottom panel ($t=1$ Myr). 
During this time the cross-like artifacts may enhance the net  
radiation energy losses. However, as their volume fraction in a single SN remnant does not exceed 3\%, their contribution to the net cooling of the  hot (postshock) 
gas  at any instant is negligible. The energy lost radiatively in such cross-like artifacts is determined by
\be 
\int dt\int\limits_{\Delta V}^{} \Lambda(T) n^2 dV\simeq n_hT_h\int dt\int\limits_{\Delta V}^{}{\Lambda(T)\over T} ndV\simlt \Lambda(T_h)n_hN_{c} 
\ee
where $n_h$ and $T_h$ is density and temperature of hot phase, $\Lambda(T)$, the cooling function, $\Delta V$ is the initial volume occupied by a growing 
cross-like structure, $N_c=\int ndV$ is the number of gas particles contained in it. In this estimate we explicitly assumed that in temperature range $T=10^{4.2}$ 
to $10^6$ K  the ratio $\Lambda(T)/T$ varies weakly, while at lower temperatures it drops because of a nearly exponential decrease of cooling function in the lower 
temperature end. Thus, as the overall cooling rate of the hot gas is proportional to $\Lambda(T_h)n_h$, the contribution of cross-like structures is proportional to $N_c$, which is proportional to
their initial volume fraction, $\Delta V$, and therefore small.

\begin{figure}
\centerline{
\epsfxsize=0.5\textwidth
\epsfbox{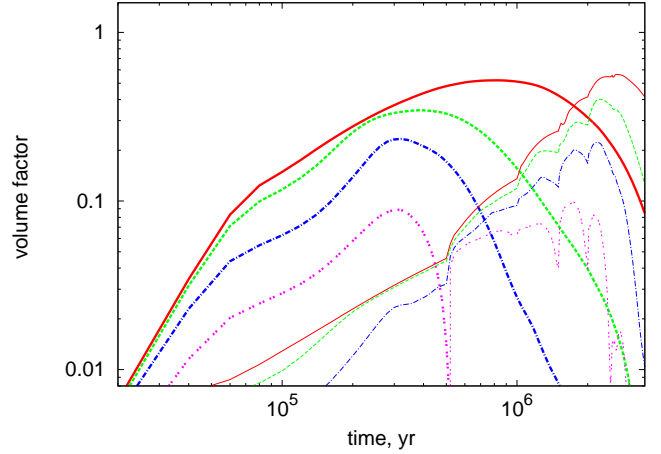}
}
{\vskip-3mm}
\caption{
Filling factors of gas with different temperatures with simultaneous SNe explosions and with
time delay. Gas number density is 1~cm$^{-3}$, with solar metallicity. 
The lines correspond to the covering factor of a gas with
$\log~T>$ 5, 6, 6.5, 7 (from top to bottom lines). 
The thick lines refer to the 15 SNe exploding in the staggered sequence of 1,2,3,4,3,2, 1 at intervals of $\Delta t=10^4$ yr, and the thin lines refer to the same sequence but with $\Delta t=5 \times 10^5$ yr.
}
\label{fig:3d-delay}
\end{figure}

 We quantify our results in terms of filling factors and heating efficiency.
Figure \ref{fig:3d-delay} shows the filling factors of gas with different
temperatures (shown in different colours), for the two cases: 
thick lines show the case of 
$\Delta t =10^4$ yr and thin lines refer to 
$\Delta t=5 \times 10^5$ yr.
The filling factors of hot gas in both cases reach similar values, albeit at different time scales. The filling factor of gas with $3 \times 10^6$ K can reach a filling factor of $0.1$ in both cases, after which it rapidly declines owing to radiative loss in merging shells.

\begin{figure}
\centerline{
\epsfxsize=0.5\textwidth
\epsfbox{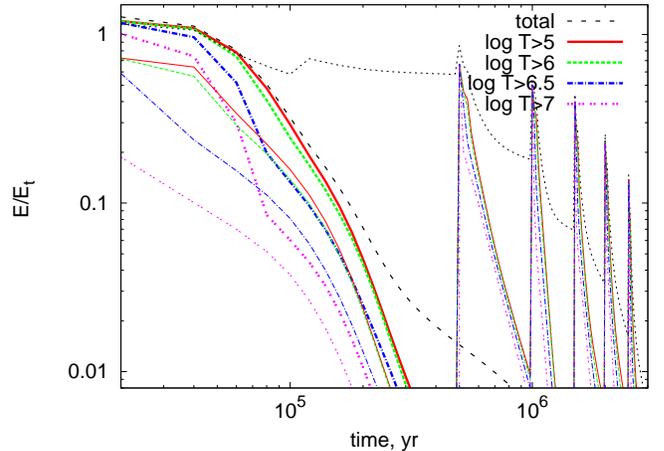}
}
{\vskip-3mm}
\caption{
The evolution of the ratio of thermal energy to the total energy, for gas in different temperature range is plotted for two different
time delays, $\Delta t=10^4$ yr (thick lines) and $\Delta t=5\times 10^5$ yr (thin lines). }
\label{fig:3deff-delay}
\end{figure}

We plot the  fractional energy of the gas at different temperatures for these cases in Figure
\ref{fig:3deff-delay}. 
For the time before the gas begins to cool precipitously, or when the explosions occur in nearly steady 
state, this fraction also gives the efficiency of SNe of heating gas up to different temperatures. 
E.g., the heating efficiency for $\Delta t=10^4$ yr  
can be estimated from the curves in the figure at $t \le 0.07$ Myr; and that for $\Delta t=5 \times 10^5$ yr 
 at a time scale of $\le 3.5$ Myr. The heating 
efficiency of gas with $T \ge 3 \times 10^6$ K lies between $\sim 0.1\hbox{--}0.3$ for the former case, and ranges between
$\sim 0.02\hbox{--}0.1$ in the latter case, with a low average value. 
Therefore, this case may have  overlapping SNe, and the filling factor of SN remnants can be as large as $\sim 0.6$, but
the heating efficiency is not large.

\subsection{Long term evolution of multiple SNe}

\begin{figure*}
\centerline{
\epsfxsize=0.85\textwidth
\epsfbox{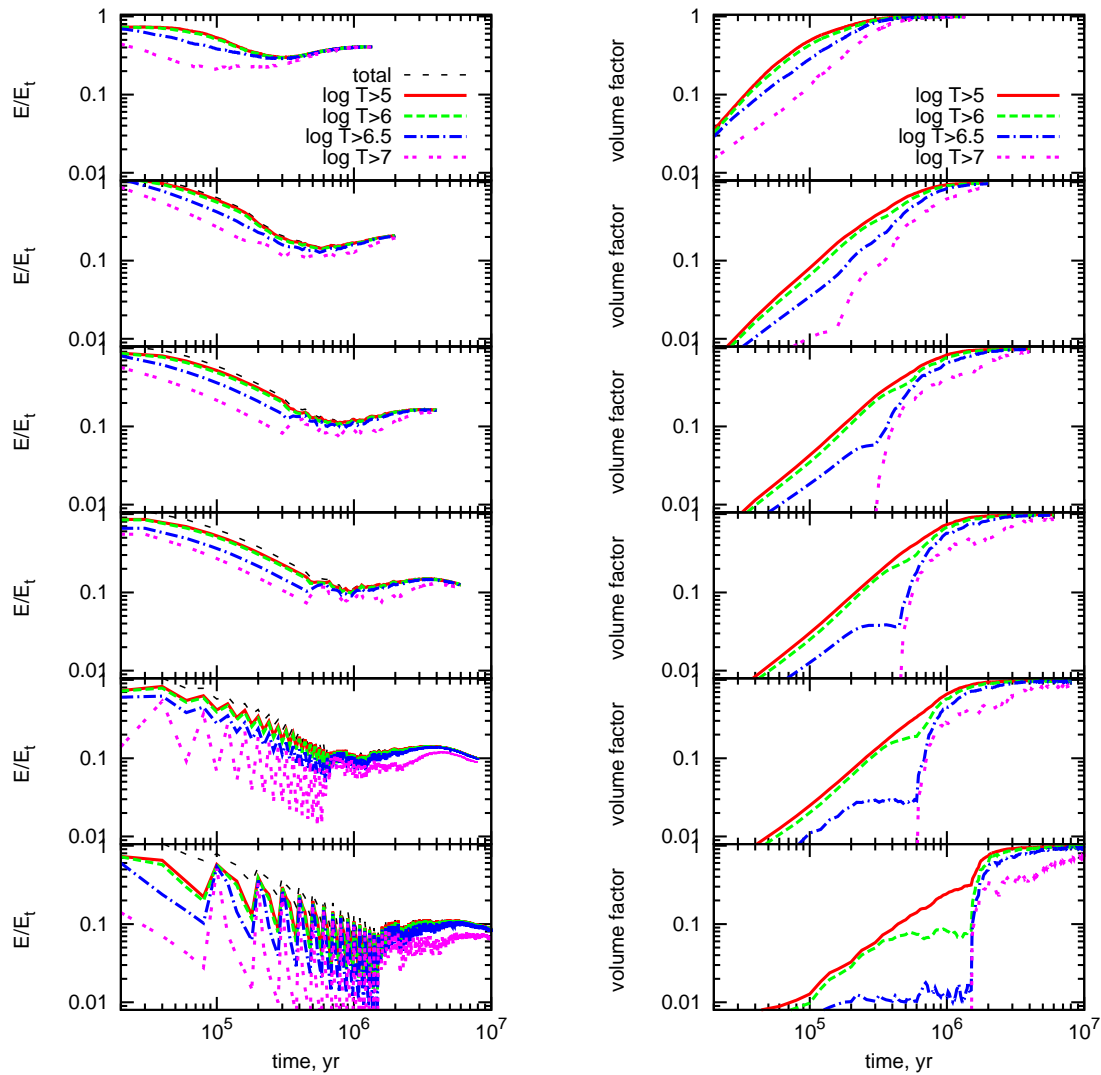}
}
{\vskip-3mm}
\caption{
The evolution of the heating efficiency (left) and filling factors (right) for gas with different temperatures, for 
a continuous series of SNe separated by time delay $\Delta=10^3, 10^4, 2 \times 10^4, 3 \times 10^4, 4 \times 10^4, 10^5$ yr (from top to bottom).
}
\label{fig:200sne}
\end{figure*}

In order to study the long term behaviour of the heating efficiency, 
we have performed runs with  SNe exploding continuously in the computational domain of 200$^3$ pc$^3$ with 
resolution of $1$ pc, with gaps of $\Delta t=10^3, 10^4, 2 \times 10^4, 3 \times 10^4, 4 \times 10^4$  
and $10^5$ yr. 
In other words, one supernova explodes  after every $\Delta t$. 
 The positions of SNe are distributed randomly in space. 
Figure \ref{fig:200sne} shows the results of heating efficiencies 
(left) and filling factors (right)  for gas with different temperatures, for all time delays (from short to long
delays, from top to bottom). We denote the efficiency of heating gas to $\ge 10^{6.5}$ K by
$\eta[10^{6.5}]$, and define it as the ratio of thermal energy stored in gas with $T \ge 10^{6.5}$ K at any given time
to the total explosion energy deposited up to that time. It is clear that the case of
more frequent SNe ($\Delta t=10^3$ yr) 
show continuous decline in the heating efficiency $\eta[10^{6.5}]$, and only after $t\simeq 10^5$ yr when the remnants 
practically fill the whole computational domain ($60\%$ of the volume), $\eta$ increases to $\sim 0.4$ because the subsequent SNe mostly expand
into hot diffuse medium. 
Explosions with a
longer delay of $\Delta t= 10^5$ yr (bottom most row) show on average similar trend on longer time scales, though as expected, with lower heating 
efficiency  of order $\eta[10^{6.5}]\sim 0.1$. Similar to the previous model, the efficiency first declines and then increases after the remnants 
 occupy roughly $30\%$ of the computational zone at $t\simeq 10^6$ yr 
to $\eta[10^{6.5}] \sim 0.1$. 

 A common feature in the behaviour of the heating efficiency in all models can be obviously noted: after a continuous decline down to $\eta(T)\simlt 0.1$, 
it stabilises and then  grows slowly for all temperature fractions, particularly for the gas with $T\geq 3\times 10^6$ K which carries a considerable 
amount of thermal energy. The most reasonable explanation is that  the epoch of increasing $\eta$  coincides with the state when the 
filling factor of the corresponding temperature fraction reaches a critical value $f(T)\sim 0.3$ when different bubbles percolate.

 This threshold filling factor can be understood in the following way, considering the evolution of isolates SN remnants.  Consider the evolution of the shell radius of an isolated SN remnant in the 
snowplough phase (valid for long term evolution
at $t> t_r$), $
R (t) \approx R_r (t/t_r)^{1/4}
$, where $R_r$ and $t_r$ are given by eqn \ref{eq:trad}. Before the widespread percolation of inner hot gas can occur, it is reasonable
to assume that most shells evolve in an isolated manner and are in the radiative phase. As far as the fraction of thermal energy in hot gas ($T\ge 10^{6.5}$ K) is concerned, one can assume that it scales as $\eta(10^{6.5})=E_{\rm th}(10^{6.5})/E\simeq 0.2 (t_r/t)$ at $t>t_r$ (see, Appendix B); here 
we neglected the enhanced radiation cooling, since the shells have not merged extensively at this phase. Suppose that the shells completely fill up the volume at time $t_c$, or that $Q (t_c)=1$. 
Since the porosity of remnants is $Q (t) =(4 \pi /3) R^3 (t) \, \nu_{\rm SN} \, t$,
we have, for $t>t_r$, $Q (t) \propto t^{7/4}$. Denoting the porosity value at the beginning of the radiative phase ($t=t_r$)
as $Q_r$, we then have $Q(t)=Q_r (t/t_r)^{7/4}$. Since by definition $Q(t_c)=1$, we have $Q_r=(t_r/t_c)^{7/4}$, at the
beginning of the radiative phase. Finally, the heating efficiency can be written as,
\be
\eta [10^{6.5}] \sim 0.2 (t_r/t_c) \sim 0.2 \, Q_r^{4/7} \,.
\ee
This shows that $Q_r \sim 0.3$ in order for the heating efficiency of hot ($T \ge 3 \times 10^6$ K) to be $\sim 0.1$.

\begin{figure*}
\centerline{
\epsfxsize=0.75\textwidth
\epsfbox{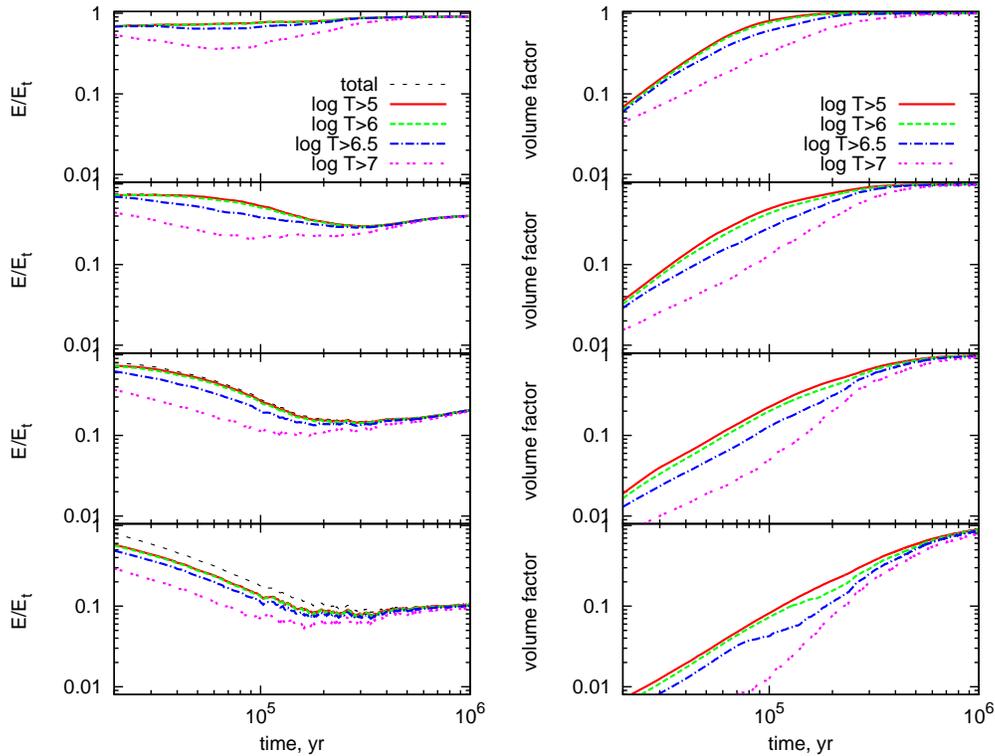}
}
{\vskip-3mm}
\caption{
The evolution of the heating efficiency (left) and filling factors (right) for gas with different temperatures, for a continuous series of SNe separated by time delay $\Delta=10^3$ yr, with different
ambient densities, $n=0.1, 1, 3, 10$ cm$^{-3}$ (from top to bottom).
}
\label{eff-den}
\end{figure*}

The time required for the percolation of hot gas can be estimated by using the result that the threshold filling factor is 
$\sim 0.3$. Consider a SNe rate density of $\nu_{\rm SN}$. Then the time required for percolation
of hot gas, $t_{\rm perc}$ can be estimated by the condition:
\be
\nu_{\rm SN} \times t_{\rm perc} \times {4 \pi \over 3} R(t_{\rm perc}) ^3 \approx 0.3\,.
\label{perc-cond}
\ee
This is however an overestimate since it assumes that all SNe explodes simultaneously at $t=0$. In reality, the total volume of remnants at $t_{\rm perc}$ is  (writing $N$ as the nearest integer to $t_{\rm perc}/\Delta t$)
\bea
{4 \pi \over 3} R_r^3 &\times&\Sigma _{n=0} ^N \Bigl ( { t_{\rm perc} -n \Delta t \over t_r} \Bigr )^{3/4} \nonumber\\
\approx {4 \pi \over 3} R_r^3 &\times&
 \Bigl ({t_{\rm perc} \over t_r} \Bigr )^{3/4} \, \Bigl [ N -{3 \over 4} {\Sigma n \over N}+ {3 \over 8} { \Sigma n^2 \over N^2}
- {3 \over 12} {\Sigma n^3 \over N^3} +\cdots \Bigr ] \,.
\eea
This introduces a factor $\sim 93/144\approx 0.65$ in the LHS of eqn \ref{perc-cond}.
For our simulations, for a given $\Delta t$, $\nu_{\rm SN} = (1/200^3) (1/\Delta t)$, and we have,
\be
t_{\rm perc} \approx 3 \, {\rm Myr} \, \Bigl ({ n \over E_{\rm 51} } \Bigr ) ^{4/7} \, \Bigl ( {\Delta t \over 10^4 \,  {\rm yr}}
\Bigr )^{4/7} \,.
\ee
The epoch of the rise of the heating efficiency in the right panels of Figure \ref{fig:200sne} is consistent with this rough estimate. We
can generalise this estimate and write,
\be t_{\rm perc}
\approx 10 \, {\rm Myr} \, \Bigl ({ n \over E_{\rm 51} } \Bigr ) ^{4/7} \, \Bigl ( {\nu_{\rm SN} \over 10^{-8} \, {\rm pc}^{-3} \, {\rm yr}^{-1}}
\Bigr )^{-4/7}
\,.
\label{tperc}
\ee
For a typical starburst SNe rate density of $\sim 10^{-9}$ pc$^{-3}$ yr$^{-1}$, and gas density of $n \sim 10$ cm$^{-3}$ in starburst
nuclei, the time scale for heating efficiency to become $\ge 0.1$ is of order $10$ Myr. 

It is interesting to note that recent observations
of 10 starburst galaxies show that there is a  time lag of $\sim 10 $ Myr between the onset of star formation and the excitation of galactic winds \citep{sharp10}. Our simulations and the important result of percolation of hot gas when the overall filling factor crosses a threshold of $\sim 0.3$, therefore, allow us to interpret this time lag as required for heating efficiency to become sufficiently large for an outflow to be launched.

 We also find that although the heating efficiency of gas increases after percolation, it decreases afterwards, and oscillates about a mean value. The reason for this behaviour is that gas keeps losing energy through radiation, but is also heated by repeated explosions.
It is interesting to note, however, that the minimum and maximum values of heating efficiency differ by a factor close to unity. Therefore
we can use the average value to infer the scaling of heating efficiency with different parameters.

For example, we can infer the scaling of the heating efficiency with the SNe rate density and ambient density from our simulations. 
We have run our simulations for different gas densities ($n=0.3, 1, 3, 10$ cm$^{-3}$) keeping the SNe frequency a constant.
Figure \ref{eff-den} shows the heating efficiencies and filling factors for these densities, with
$\Delta t=10^3$ yr.
We find that roughly $\eta[10^{6.5}] \propto \nu_{\rm SN}^{0.2} n^{-0.6}$, for the heating efficiency of X-ray emitting gas.
For our case of $\Delta t=10^3$ yr, and $n=1$ cm$^{-3}$, the SNe rate density 
corresponds to  $\sim  10^{-10}$  pc$^{-3}$ yr$^{-1}$, and $\eta[10^{6.5}]\approx 0.35$. 
Therefore, for a typical SNe rate density in starburst nuclei of 
$\sim 10^{-9}$  pc$^{-3}$ yr$^{-1}$, and gas density $n\approx 10$ cm$^{-3}$, the heating efficiency is  $\eta[6.5]\approx 0.15$.

\section{Discussion}
Our aim in this paper has been to perform controlled numerical experiments in order to study the effect of coherency and long term effects of SNe events, instead of linking the SNe events to the underlying star formation process. Our focus has been on the filling factor and fraction of energy stored in the hot gas at different temperatures, especially for the X-ray emitting gas with $T\ge 3 \times 10^6$ K.

One of our main results is that the radiative loss in the case of multiple SNe events can be {\it much larger} than in the 
case of single SN remnants. In comparison to the loss rate of total thermal energy, scaling roughly as $t^{-1}$ 
\citep{chevalier74b, cox72, thornton98}, the loss for hot gas 
scales steeply with time, 
as $t^{-2.3}$ for gas with $T \ge 10^6$ K, which is due to an enhanced radiation in multiple overlapping shocks. 
This result has important implications for the heating efficiency of SNe and the filling factors of hot gas.

Our results show that the filling factors crucially depend on the comparison between the radiative time scale $t_r$ and the collision time scale of SNe shells, $t_c$. Moreover, the SNe should continue beyond the time scale for percolation of hot gas.
 The fact
that $t_r$ is an important time scale has been recognized since the seminal work by \citet{larson74}. However it has been tacitly assumed that the filling factor of $T\ge 10^6$ K is of order unity \citep{heckman90}. Recently \citet{nath13} have pointed out that this assumption is not valid in the conditions prevailing in the central regions of starbursts, because of prohibitive radiative loss in large ambient density. This criticism is, by the way, not in contradiction with the assumption \citep{strickland09} that the heating efficiency can be large, of order $0.1$. The question is whether or not the filling factor of X-ray emitting gas in the central ($200\hbox{--}300$ pc) of starbursts can be regarded as close to unity, as is required in the galactic wind models of \citet{clegg85, sharma13}.

Our result that the heating efficiency 
can be  $\sim 0.1\hbox{--}0.2$ for typical starburst nuclei region parameters,
is consistent with the values inferred from X-ray observations of starburst driven outflows. 
\citet{strickland09} have inferred a value of $0.1\hbox{--}0.3$ for the heating efficiency of SNe for the X-ray emitting gas. 
Our results put these inference and assumptions on a firm footing, and also provide scalings with the SNe
rate density and gas density so that heating efficiency can be estimated for a general case.
Moreover, the requirement for percolation of hot gas provides a natural explanation for the observed time lag between
the onset of star formation and the launching of galactic winds, as mentioned earlier.

\begin{figure*}
\includegraphics[width=9.5cm, angle=270]{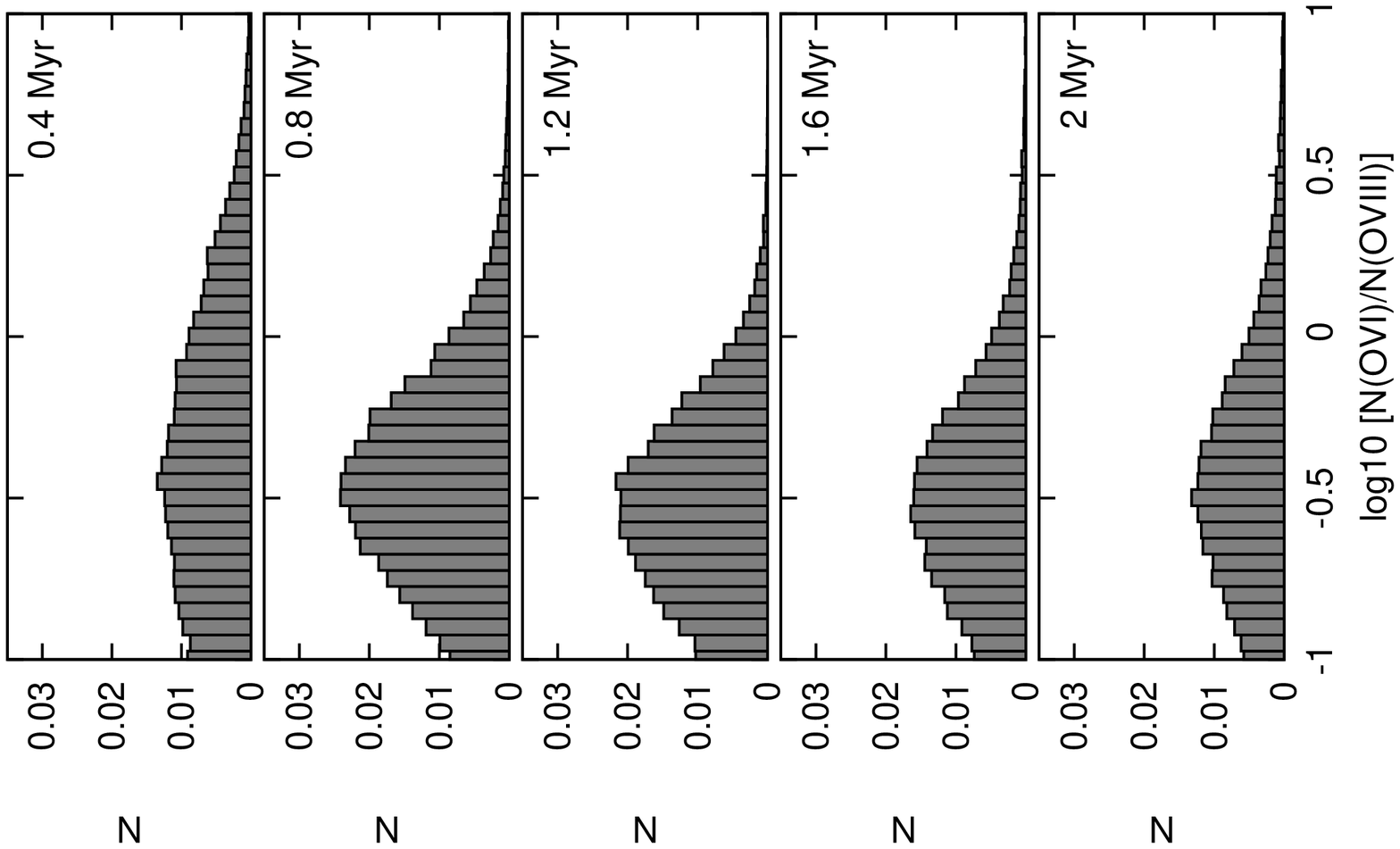}
\includegraphics[width=9.5cm, angle=270]{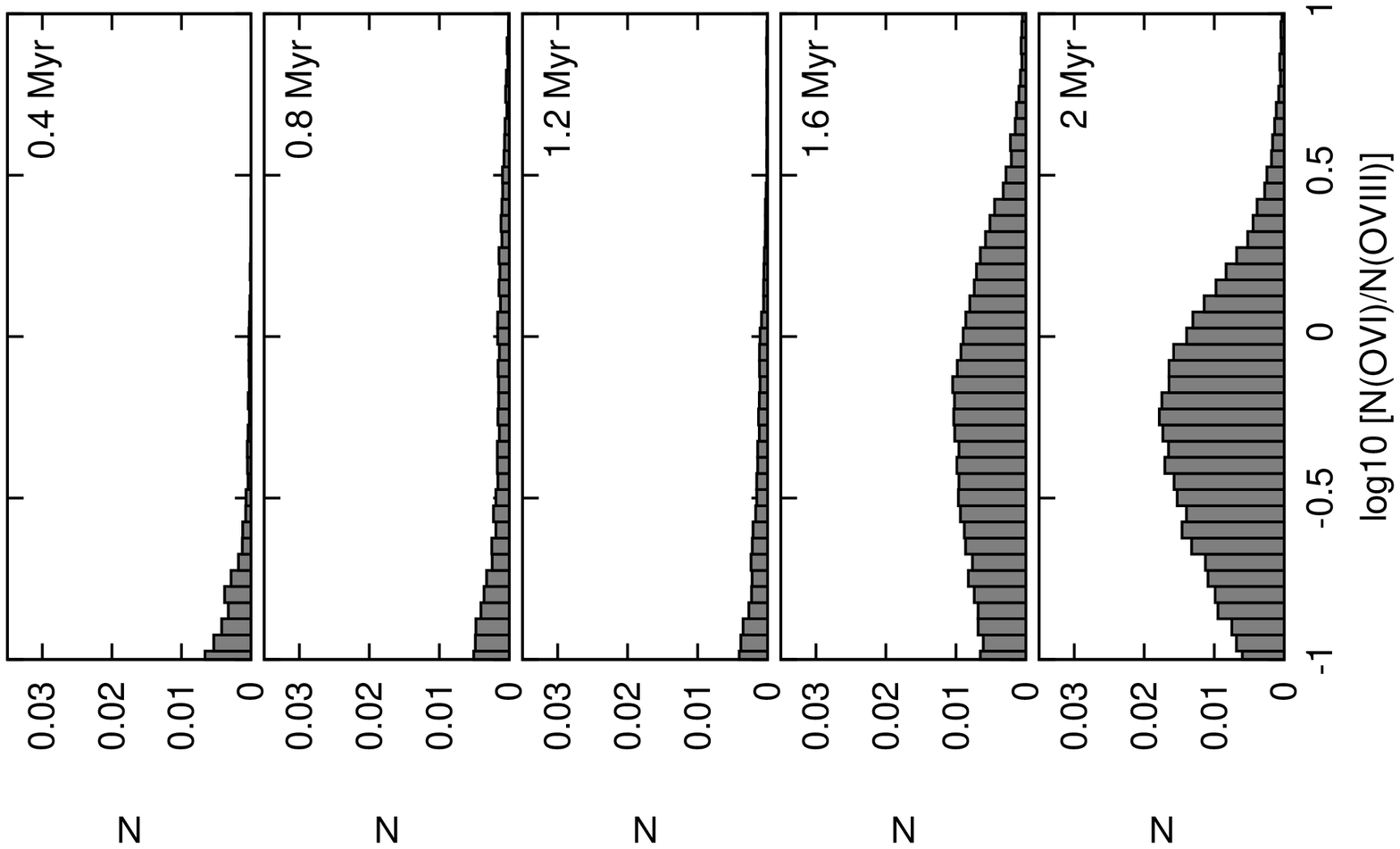}
\caption{
The distributions of the ratio of OVI to OVIII column densities at a few snapshots (0.4, 0.8, 1.2, 1.6, 2 Myr), for two different time delays between the SNe, with $\Delta t=10^4$ yr on the left, and $\Delta t=10^5$ yr on the right.
}
\label{fig:hist0}
\end{figure*}

 It is clear from our simulations that in order to excite galactic winds it is crucial that old SNRs are not allowed to cool to low temperatures (below about $ 10^6$ K). Since different ions can probe different temperatures, one can use the abundance ratios of relevant ions to probe the nature of the hot gas in starbursts. Observations, such as that of filamentary structure of H$\alpha$ emission in a multiple SNe environment by \citep{egorov14}, could be important in this regard.

For example, the ions OVI and OVIII probe gas with temperatures differing  by an order of magnitude, of $10^{5.5}$ K and $10^{6.5}$ K respectively. The abundance ratios of these two ions can therefore shed light on the filling factor  of gas in these temperature ranges. 
It is instructive to divide the ionic ratio in three main parts: (a) when $N({\rm OVI})/N({\rm OVIII}) \ll 1$, young SNRs dominate the region, with fast shocks and high temperature gas; (b) when $N({\rm OVI})/N({\rm OVIII}) \sim 1$, SNRs enter the radiative phase, and (c) when the ratio is much larger than unity, SNRs become old and warm instead of being hot.
Since the effect of SNe in the case of coherency is not to allow SNRs to cool significantly, the
coherency condition should manifest as the ratio $N({\rm OVI})/N({\rm OVIII})$ being in the range $\sim 0.1\hbox{--}1$, and not exceed of order unity.

We plot the distribution of the OVI to OVIII column density ratio for a few snapshots of time in
Figure \ref{fig:hist0} for $\Delta t=10^4$ yr (left panel)  and $\Delta t=10^5$ yr (right panel).  The former case refers to 
a larger SNe rate density.
The histograms show that for the case of 
smaller SNe rate density, the distribution is biased towards smaller
values of OVI to OVIII ratio, whereas for the 
case  larger SNe rate density, there are regions with $N({\rm OVI})/N({\rm OVIII}) <1$ as well as with $N({\rm OVI})/N({\rm OVIII})>1$, with an average near unit ratio.  The histograms become similar when the percolation of hot gas is over (bottom most
panels). This is expected because with decreasing SNe rate density, the interval between SNRs is long, and low temperature
gas dominates due to cooling, which decreases the ratio of $N({\rm OVI})/N({\rm OVIII})$, until the hot gas has a chance to percolate.
The ratio of these two column densities can therefore be a diagnostic of the effect of multiple SNe.

\begin{figure*}
\includegraphics[width=9.5cm, angle=270]{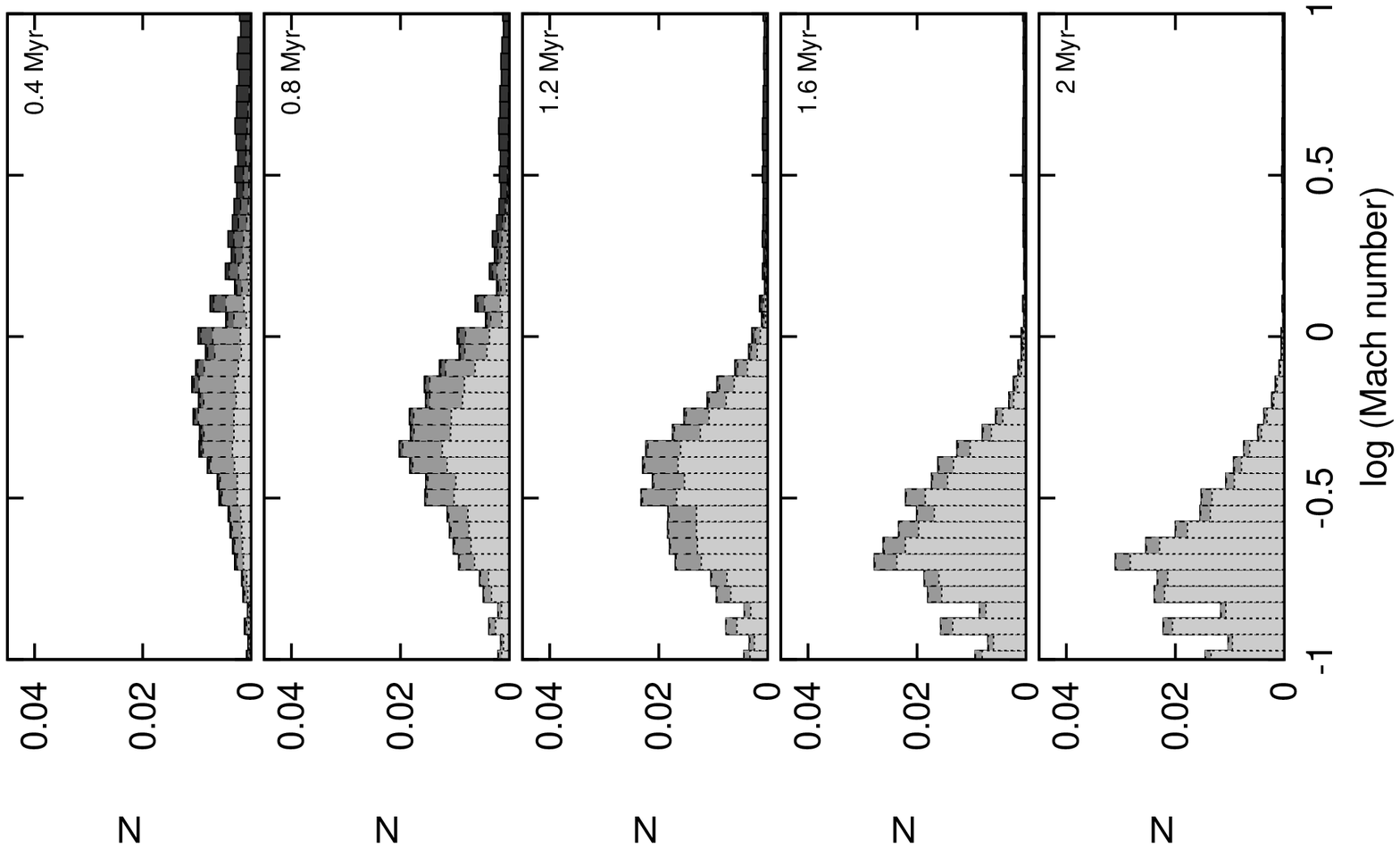}
\includegraphics[width=9.5cm, angle=270]{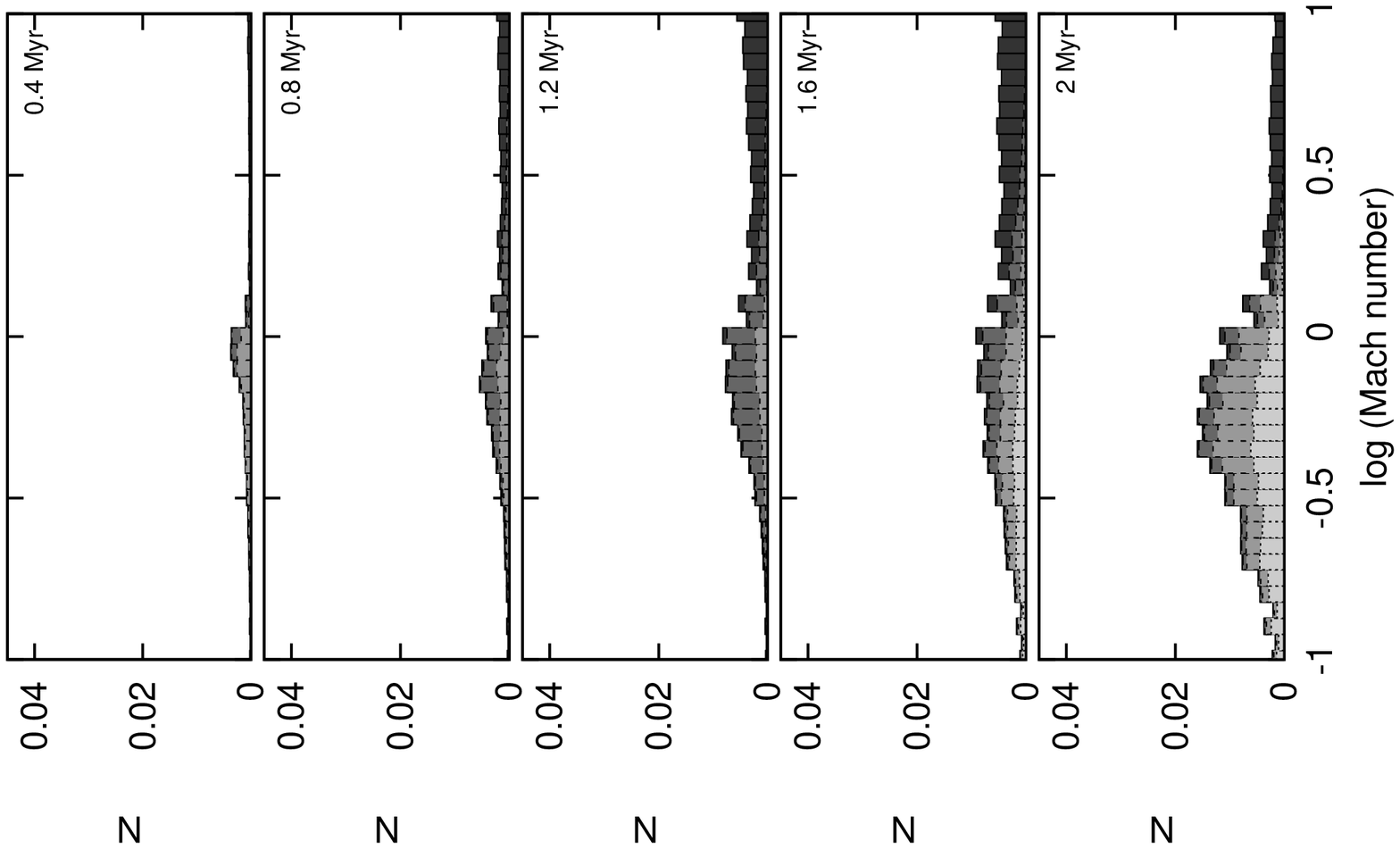}
\caption{
The distribution of Mach numbers of gas with different temperatures at different epochs, with two different time delays, $\Delta t=10^4$ yr on the left, and $\Delta t=10^5$ yr on the right. Three temperature bins are considered here: $T>10^5$ K (dark), $T > 10^6$ K (grey) and $T > 10^7$ K (light grey).
}
\label{fig:hist-mach}
\end{figure*}

Finally, we show the distribution of Mach numbers of gas at three different temperature bins in Figure \ref{fig:hist-mach}, for two different time delays. We consider gas in three temperature bins: $\log T > 5, 6, 7$, which are shown with dark, grey and light grey shades, respectively. In other words, hotter phases of gas are shown with relatively lighter shades. The distributions of Mach numbers show that the hottest gas has Mach number less than unity (subsonic) on average, although with a scatter.  In other words, the width of the emission lines from the highly ionized species from the hottest parts of gas would be dominated by thermal spread. In contrast, the Mach numbers of relatively colder gas have a large scatter and reach high values ($\le 3$). This implies that the width of  emission or absorption lines from low ionization species can be dominated by non-thermal, turbulent motions. In other words, the high and low ionization species are likely to trace different dynamical states. This is also reasonable because the post-shock gas is mostly subsonic.

\section{Summary}
We have studied the effect of multiple SNe on the filling factor of hot gas and the efficiency of heating gas up to high temperatures. We have tested the idea that the filling factor of hot gas and the heating efficiency depend strongly on whether or not the SNe explosions satisfy the coherency condition, which can expressed succinctly as $t_c \le t_r$ (time scale of SNR overlap being smaller than the radiative loss time scale).

Our 
3D hydrodynamical simulations show that this is indeed true, and that radiative cooling is more pronounced in the case of multiple SNe events than in single SN remnants. While in the case of single SN remnant, the thermal energy drops as $t^{-0.6}$ after the radiative phase, in the case of multiple (simultaneous) SNe, the total thermal energy scales as $t^{-1.5}$ (and, for hot gas the energy scales even more steeply as $t^{-3.5}$) , owing to large densities in merging shells and consequent enhanced cooling. This has significant implications for the filling factors of hot gas and the overall heating efficiency of multiple SNe.

Our simulations show that in the case of continuous series of SNe, hot gas can percolate throughout the region of star formation,
after a time $\sim 10$ Myr, for typical starburst nuclei parameters. This is consistent with observations of \citet{sharp10} who found a 
time lag of a similar order between the onset of star formation and the launch of a galactic wind. We determined the efficiency of 
heating the gas to X-ray temperatures ($\ge 10^{6.5}$ K) to be $\sim 0.1\hbox{--}0.2$ for typical SNe rate density ($\nu_{\rm SN}\approx 10^{-9}$ pc$^{-3}$ yr$^{-1}$) and gas density ($n\approx 10$ cm$^{-3}$) in starburst nuclei. Our simulation shows that the heating
efficiency scales as $\eta \propto \nu_{\rm SN}^{0.2} \, n^{-0.6}$, which can be used to estimate the heating efficiency in other cases.
 
 Based on our simulations, we have suggested that the ionic ratio of OVI and OVIII could reveal the effect of SNe. 
 We found that before the percolation of hot gas occurs, the column density ratio of OVI to OVIII is unlikely to exceed unity. We have also suggested that the widths of emission/absorption lines from hot and warm gas are dominated by thermal and turbulent motions. These can be used as a diagnostic of the physical state of starburst regions.

\bigskip
We thank Biswajit Paul, Prateek Sharma and Daniel Wang for helpful discussions and the anonymous referee for valuable suggestions.
This work is partly supported by an Indo-Russian project (RFBR project code 12-02-92704, DST-India project code INT-
         RFBR-P121). EV and YuS thank to the RFBR for support (project codes 12-02-00365 and 12-02-00917).
         EV is grateful for support from the Dynasty Foundation.

\appendix

\section{Cooling rates}
In metal enriched collisional gas with $Z\simgt 0.1~\zsun$ the cooling by metals becomes dominant at $T\simlt 10^7$~K 
[see e.g. \citep{wiersma,v11}]. 
When the cooling time becomes shorter than the age of a SN remnant, the shell starts to lose much energy through 
radiative processes. Until the radiative phase begins, the ionization/recombination processes in SN shell are fast, 
their time scales are shorter than the age of the remnant. Therefore, the ionic composition of a gas is in collisional 
equilibrium. However, when radiative losses become significant the equilibrium is broken, and the temperature decreases
faster than the ionic composition is able to settle into an equilibrium. This originates from the fact that 
the recombination time scales of metal ions become longer than the cooling time. Therefore the gas remains 
over-ionized in comparison with that in an equilibrium state. Consequently, the equilibrium cooling rates cannot 
be applicable for SN shell at radiative phase. One should use the non-equilibrium (time-dependent) cooling rates 
for studying SN shell evolution.

The self-consistent calculation of cooling rates in multi-dimensional dynamics of SN shell is a 
time consuming task. However the evolution of gas behind shock waves with velocities higher than $\simgt 
150$~km~s$^{-1}$ are very close to that of a gas cooled from very high temperature $T=10^8$~K \citep{v12, sd93}
[see their Fig.11]. 
Therefore the non-equilibrium cooling rates can be pre-computed for a gas cooled from very high temperature, 
e.g. $T=10^8$~K and these rates can be used to study SN shell evolution in tabulated form. This approach 
is applicable for a gas with $T\sim 10^4$~K that passes through the shock front with velocity higher than $\simgt
150$~km~s$^{-1}$. 
Of course, 
if a parcel of neutral gas passes through the shock front, some additional time is needed for ionization, i.e.,
relaxation. This relaxation time scale may be long enough, but during this period the ionic composition tends 
to the non-equilibrium values in the collisional case \citep{v12}. 

In this paper we study multiple SN explosions, whose shells collide and merge with each other during the evolution. 
The typical velocities of colliding shells 
are higher than 100~km~s$^{-1}$. Even if some 
parts of the shells have smaller velocity, they eventually collide with one another with such a shock that the relative 
velocity of gas flows becomes more than 100~km~s$^{-1}$ and our non-equilibrium cooling rates in tabulated form 
are applied.

\begin{figure}
\includegraphics[width=80mm]{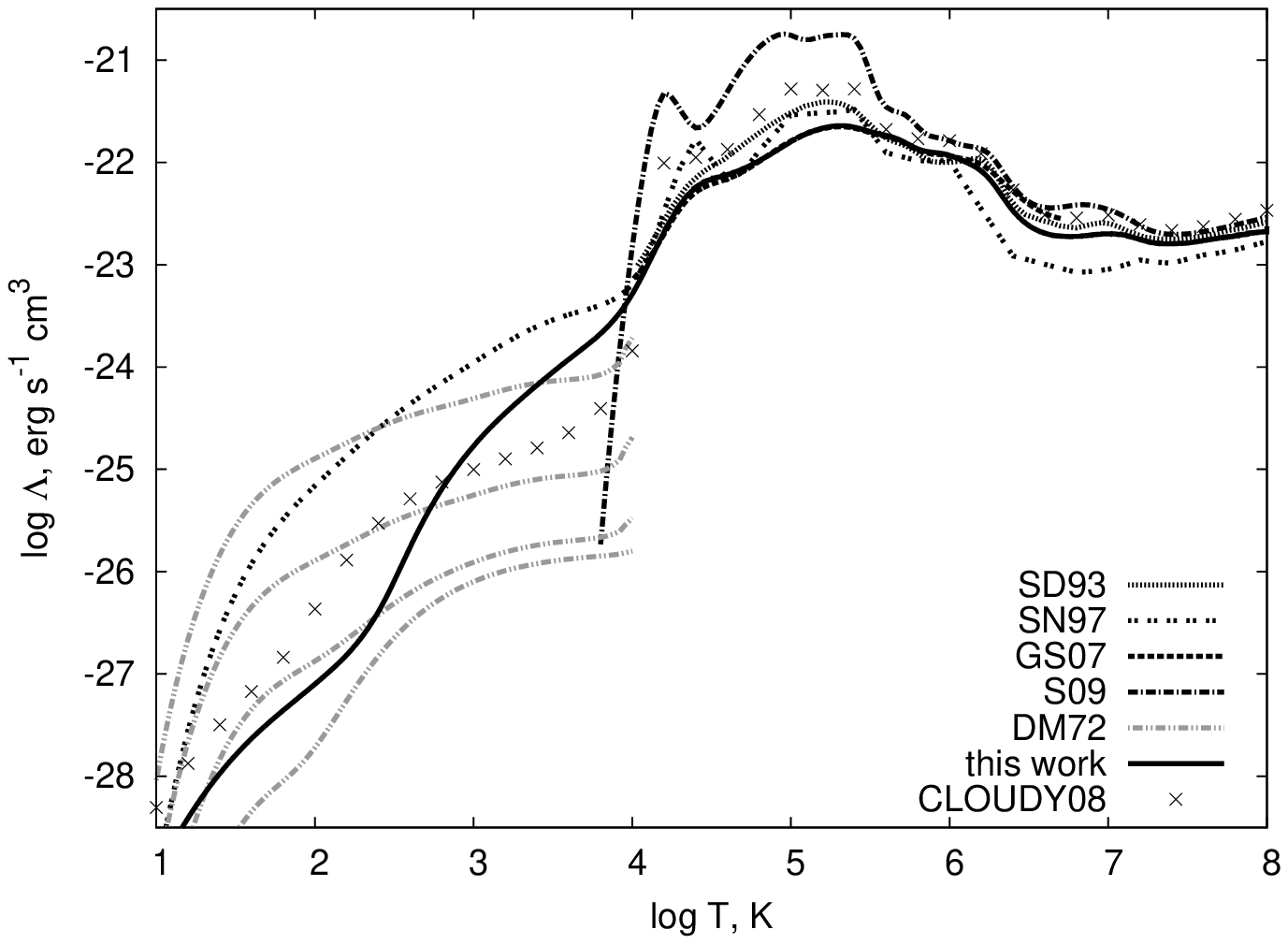}
\caption{
The cooling rates for solar metallicity. The isobaric rate for gas with initial density $n=1$~cm$^{-3}$
calculated by \citet{v13} and used in this work is depicted by solid line. The other lines correspond 
to the rates obtained in the previous calculations: the cooling rates obtained by \citet{sd93} (SD93, 
we have chosen their non-equilibrium data), \citet{sn97} (SN97), \citet{gs07} (GS07), \citet{schure09} 
(S09), and \citet{dalgarno72} for ionization fraction, $f_i=n_e/n_H = 10^{-4}$, $10^{-3}$, $10^{-2}$, 
$0.1$, which are shown by dash-dot-dotted lines from bottom to top. The data obtained from {\small CLOUDY} 
code \citep[v.08,][]{cloudy} is depicted by crosses (the H$_2$ molecules are ignored in the equilibrium calculation).
}
\label{figcoolsol-example}
\end{figure}

Non-equilibrium cooling rates are generally lower than the equilibrium rates at $T\simlt 10^6$~K
(Figure~\ref{figcoolsol-example}), whereas for higher temperature the rates are close to each other. 
The 
radiative phase commences in SN shell at $T\sim 10^6$~K, almost independent of which rates are used. Further evolution
tracks of the shell certain differ. Because of higher equilibrium cooling rates the shell cools faster down to $T\sim
10^4$~K. At this temperature the use of equilibrium rates faces at least two problems. Firstly, few data of cooling
rates for $T\simlt 10^4$~K can be found. The most commonly used rates are calculated by \citet{dalgarno72}. Secondly,
these rates are obtained by summing the rates of main cooling agents, and they depend on one (or more) free parameter. 
For example, to use the \citet{dalgarno72} cooling rates one should manually put
the ionization fraction. This free parameter is set constant for a gas in the whole temperature range $T\simlt 10^4$~K, whereas self-consistent calculations show a significant dependence on the temperature and demonstrate 
qualitative difference for gas with different metallicity \citep[see Figures 4 and 6 in][]{v13}. The lack
of cooling at $T\simlt 10^4$~K in the case of equilibrium cooling rate with fixed ionization fraction leads
to the formation of thick shell with more or less constant temperature \citep[see Figure 9 in][]{v13}. The thickness
of this shell depends on the manually chosen ionization fraction. It exists for a significant time compared to the 
age of the shell. Certainly, such thick shells can change the dynamics of SN shell collisions, higher equilibrium 
cooling rates lead to more effective energy losses in the shell during radiative phase. These inconsistencies allow
us to use non-equilibrium cooling rates with better confidence.

The full description of our method of cooling rate calculations and the references to the atomic data can be found 
in \citet{v11,v13}. Briefly, the chemical and thermal evolution of a gas parcel can be divided into high-temperature 
($T> 2\times 10^4$~K) and  low-temperature ($T\le 2\times 10^4$~K) ranges. Such division is motivated by the
transition to neutral gas and the formation of molecules in the latter range. In the former we consider all ionization
states of the elements H, He, C, N, O, Ne, Mg, Si and Fe. We take into account the following major processes in a
collisional gas: collisional ionization, radiative and dielectronic recombination as well as charge transfer in 
collisions with hydrogen and helium atoms and ions. In order to calculate the rates in a low-temperature ($T\le 2\times
10^4$~K) gas, the above listed ionization states of the elements are supplemented by a standard set of species: 
H$^-$, H$_2$, H$_2^+$, D, D$^+$, D$^-$, HD, needed to model the H$_2$/HD gas-phase kinetics \citep{abel97,galli98}. 

\section{Convergence tests}
\begin{figure}
\centerline{
\epsfxsize=0.5\textwidth
\epsfbox{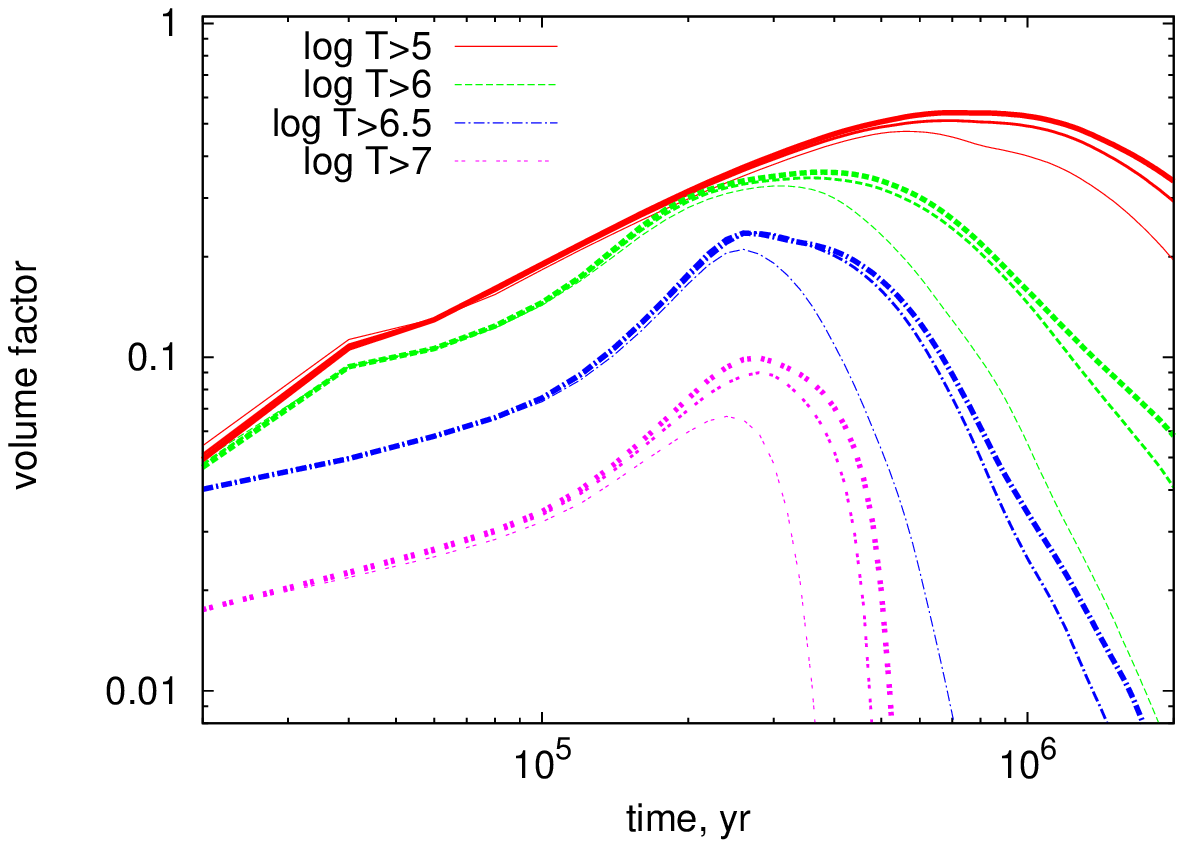}
}
{\vskip-3mm}
\caption{
Convergence test for the case of simultaneous explosions. Filling factors of gas with different temperatures for simultaneous SN explosions are shown for runs with different resolutions ($200^3$ (thinest lines), $300^3$ (thicker lines) and $400^3$ cells (thickest lines), from bottom to top). Gas number density is $1$ cm$^{-3}$, with solar metallicity.}
\label{fig:3dconv}
\end{figure}

In order to test the convergence of our simulations, we have re-simulated the model with simultaneous SN explosions with different resolutions, with computational domains of size $200^3$,  $300^3$ and $400^3$ cells.
Figure \ref{fig:3dconv} shows the filling factors for gas with different temperature for runs with different resolutions. It is seen that runs with
$300^3$ cells (the results of which are discussed here) are close to the convergence limit.

\section{Isolated explosion} 
\begin{figure}
\includegraphics[width=80mm]{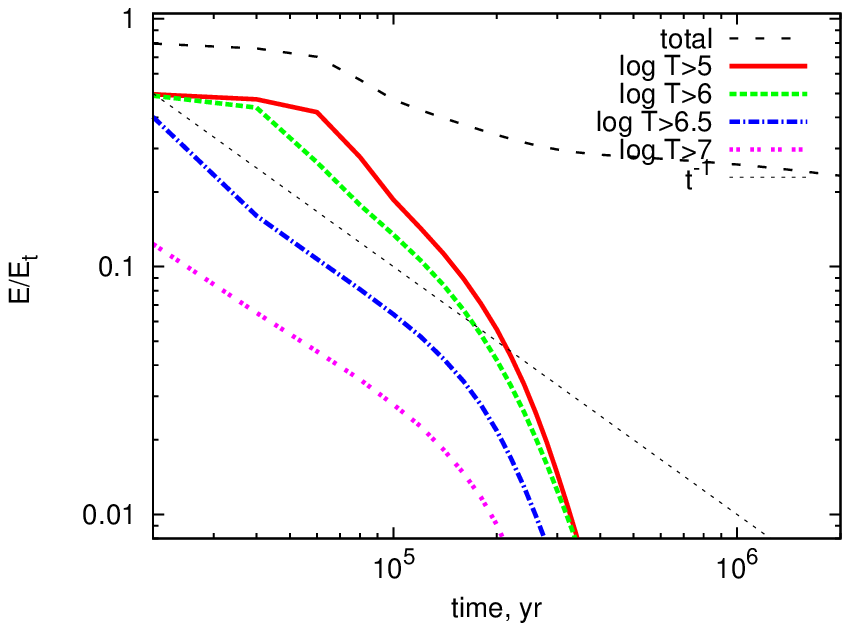}
\caption{
The evolution of the heating efficiency -- energy fraction stored in different temperature bins for 
isolated SN explosion. 
}
\label{esingle}
\end{figure}

In order to understand how the energy fraction is stored in different temperature bins when multiple 
SNe remnants overlap, here we show the evolution of a single SN remnant. The explosion is treated in the 3-D model 
with the standard energy explosion $E=10^{51}$ erg in a medium with $n=1$ cm$^{-3}$, and with resolution $\Delta x=1$ pc. 
In Figure \ref{esingle} we show how $\eta(T)=E_t(T)/E$ changes with time. Hot gas with $T>10^{6.5}$ K begins to cool 
immediately after the shock enters the radiative phase $t>t_r$ (here $t_r=2 \times 10^4$ yr). A comparison with the 
dashed line with a slope of 
$t^{-1}$ shows that $\eta \propto t^{-1}$, before enhanced radiation loss makes the slope steeper. For a limited period
of time, we can approximate it as $\eta=0.2 (t/t_r)^{-1}$.

\end{document}